\documentclass{jaa}

\usepackage{graphicx}

\begin{document}

\title{Modified Newtonian Dynamics (MOND) as a Modification of Newtonian Inertia}


\author{Mohammed Alzain}
\affilOne{Department of Physics, Omdurman Islamic University, Sudan.}


\twocolumn[{

\maketitle

\corres{malzain1992@gmail.com}

\msinfo{}{}{}

\begin{abstract}
 We present a modified inertia formulation of Modified Newtonian dynamics (MOND) without retaining Galilean invariance. Assuming that the existence of a universal upper bound, predicted by MOND, to the acceleration produced by a dark halo is equivalent to a violation of the hypothesis of locality (which states that an accelerated observer is pointwise inertial), we demonstrate that Milgrom's law is invariant under a new space-time coordinate transformation. In light of the new coordinate symmetry, we address the deficiency of MOND in resolving the mass discrepancy problem in clusters of galaxies.
\end{abstract}


}]


\doinum{}
\artcitid{}
\volnum{}
\year{2017}
\pgrange{1--9}
\setcounter{page}{1}
\lp{9}

\section{Introduction}

The modified Newtonian dynamics (MOND) paradigm posits that the observations attributed to the presence of dark matter can be explained and empirically unified as a modification of Newtonian dynamics when the gravitational acceleration falls below a constant value of $a_0\simeq 10^{-10} ms^{-2}$. Milgrom \cite{Milgrom1} noticed that the rotation curves of disk galaxies can be specified given only the distribution of visible (baryonic) matter, using the formula
\begin{equation}\label{eq:mil0}
g\mu (g/a_0)=g_N
\end{equation}
which relates the observed gravitational acceleration $g$ to the Newtonian gravitational acceleration $g_N$ as calculated from the baryonic mass distribution. The interpolating function $\mu (g/a_0)$ satisfies $\mu (g/a_0)=g/a_0$ when $g\ll a_0$, and $\mu (g/a_0)=1$ when $g\gg a_0$. The direct observational evidence for Milgrom's formula is the fact that the mass discrepancy in galaxies of all sizes always appears below the acceleration scale $a_0$ \cite{Famaey}. It follows from the appearance of an acceleration scale where dark matter halos are needed that there is a universal upper bound to the acceleration that a dark halo can produce \cite{brada}. The difference between the MOND acceleration $g$ and the Newtonian acceleration $g_N$ can be explained by the presence of a fictitious dark halo and the upper bound is inferred by writing the excess (halo) acceleration as a function of the MOND acceleration,
\begin{equation}\label{eq:halo}
g_D(g)=g-g_N=g-g\mu (g/a_0).
\end{equation}
It seems from the behavior of the interpolating function as dictated by Milgrom's formula \cite{brada} that the acceleration Eq. (\ref{eq:halo}) is universally bounded from the above by a value of order $a_0$,
\begin{equation}\label{eq:max}
a_{\dag} =\eta a_0,
\end{equation}
where $\eta $ is a dimensionless constant which is of order unity. This prediction was confirmed from the rotation curves for a sample of disk galaxies \cite{sanders}.

On the other hand, in Einstein's special theory of relativity, when Lorentz invariance is extended to accelerated observers it is assumed that the behavior of measuring rods and clocks is independent of acceleration \cite{Mashhoon}. This in fact, is a statement of the hypothesis of locality which asserts that an accelerated observer makes the same measurements as a hypothetical momentarily co-moving inertial observer. For instance, the rate of an accelerated clock is assumed to be independent of its acceleration and identical to that of the instantaneously co-moving inertial clock ``the clock hypothesis''\cite{Moller, Rindler}.

If, however, we assume that the characteristic maximum acceleration that appears in the behavior of dark halos predicted by MOND Eq. (\ref{eq:max}) is invariant under transformations from inertial to accelerated reference frames, then our assumption apparently contradicts the kinematic rule that the acceleration $a_{\dag}$ as measured in an inertial frame $S$ is given by \cite{French}: $a_{\dag}=a_{\dag}'+A$, where $a_{\dag}'$ is the acceleration as measured in an accelerated frame $S'$ and $A$ is the acceleration of the frame $S'$ with respect to the inertial frame $S$. Thus, if we assume that $a_{\dag}=a_{\dag}'$, then the accelerated measuring rods and clocks must behave in such a way that the relative acceleration between the two reference frames $S'$ and $S$ becomes undetectable and therefore it becomes unreasonable to assume that the hypothesis of locality is still valid upon making such assumption about the maximum halo acceleration. So, probably the most suitable way to derive the maximum halo acceleration Eq. (\ref{eq:max}) from physical assumptions (not from the near coincidence of $a_0$ with cosmological parameters \cite{Milgrom1} or introducing any new assumptions about the nature of dark matter) is to assume that the hypothesis of locality is false in the low acceleration limit $g\ll a_0$.

Our derivation of the maximum halo acceleration is based on interpreting Milgrom's formula as a modification of Newton's second law of motion \cite{Milgrom},
\begin{equation}\label{eq:mil}
F=mg\mu (g/a_0),
\end{equation}
where $F$ is the total force exerted on the particle, and $m$ is the inertial mass (response of the particle to all forces). However, this law is not enough by itself to represent a consistent modification of Newtonian inertia, because if we consider an isolated system consisting of two bodies interacting gravitationally with small masses $m_1$ and $m_2$ such that Eq. (\ref{eq:mil}) applies, and differentiate the total momentum $p=p_1+p_2$ using Eq. (\ref{eq:mil}) we obtain \cite{Felten}
\begin{equation}\label{eq:mcon}
\frac{dp}{dt}=\sqrt{a_0 F}(\sqrt{m_1}-\sqrt{m_2}).
\end{equation}
The total momentum of this isolated system is not conserved $\frac{dp}{dt}\neq 0$ unless $m_1=m_2$, this problem can be avoided if there is a nonstandard kinetic action from which the equation of motion Eq. (\ref{eq:mil}) is derived. Milgrom \cite{Milgrom} constructed such modified kinetic actions and showed that they must be time-nonlocal to be Galilean invariant, but it is unclear how to construct a relativistic generalization of such scheme. It should be mentioned here that there are other possible approaches to MOND inertia, for example, inertia could be the product of the interaction of an accelerating particle with the vacuum \cite{Milgrom3, McCulloch}.

The problems of modified inertia formulations of MOND can be alleviated by interpreting Milgrom's formula as a modification of Newtonian gravity. Bekenstein and Milgrom \cite{bekenstein2} proposed a non-relativistic theory of MOND as a modification of Newtonian gravity (called AQUAL); the theory contains a modified gravitational action while the kinetic action takes its standard form, and thus conservation laws are preserved. The attempts to formulate a covariant generalization of AQUAL culminated with the emergence of the Tensor-Vector-Scalar theory (TeVeS) \cite{bekenstein} the first consistent relativistic gravitational field theory for MOND, but even in this theory there is still a need for a predefined interpolating function that interpolates between the Newtonian and MONDian regime.

Instead of focusing our attention on constructing modified actions for MOND, let us reconsider the non-conservation of momentum problem from a mathematical point of view; the non-conservation of momentum exhibited in Eq. (\ref{eq:mcon}) can be attributed to the fact that each body in the isolated system is subject to a non-Newtonian force of magnitude $ma_0$, thus
\begin{equation}
\frac{dp}{dt}=\sqrt{F}(\sqrt{m_1 a_0}-\sqrt{m_2 a_0}),
\end{equation}
which means that it is not possible to isolate the two interacting bodies in the low acceleration limit from the influence of all sorts of external forces. So, why should we expect Newton's first law to be valid in the MOND regime? perhaps, the isolated body in the MOND regime behaves differently from the isolated body in the Newtonian regime.

Newton's first law states that for an isolated body, far removed from all other matter, the vector sum of all forces vanishes $\vec{F}=0$, hence, the isolated body moves with uniform velocity. Let us assume instead that in the case of an isolated body in the MOND regime; the sum of the magnitudes of all forces is constant and is proportional to $ma_0$, thus, $F=\eta ma_0$ where $\eta $ is the proportionality factor, $a_0$ is the MOND acceleration constant, and $m$ is the inertial mass. The isolated body then moves with uniform acceleration $g\mu (g/a_0)=\eta a_0$, we will refer to this assertion as the modified Newton's first law. Note that this is an assertion that cannot be confirmed experimentally, like Newton's first law.

Suppose an observer is placed in a reference frame $S$ in which the modified Newton's first law holds, then the observer in this frame will measure a force $F=mg\mu (g/a_0)=m\eta a_0$. If there is another observer in a frame $S'$ which is moving with respect to $S$ with acceleration, then the second observer will also measure the same force $F'=mg'\mu (g'/a_0)=m\eta a_0$ (assuming that the mass and the acceleration constant $\eta a_0$ are the same in $S$ as in $S'$). The relative acceleration between the two frames $S$ and $S'$ is not dynamically detectable due to the invariance of $\eta a_0$.

Therefore, the modified Newton's first law, coupled with the invariance of $\eta a_0$, defines an infinite class of equivalent reference frames in accelerated motion relative to one another and suppresses the appearance of inertial forces, hence, the uniformly accelerated frames $S$ and $S'$ are equivalent and Milgrom's law is the same in both frames $F=mg\mu (g/a_0)=mg'\mu (g'/a_0)=F'$. When a MOND theory is fully compatible with this coordinate symmetry (i.e. the impossibility of detecting a coordinate change) it must satisfy conservation laws such as the conservation of momentum, because the coordinate symmetry implies the homogeneity of space. The modified Newton's first law is a key feature of our derivation of the upper bound Eq. (\ref{eq:max}) from physical assumptions.

\section{The maximum halo acceleration and its consequences}

Any physical process that involves the dynamics of particles and fields plays out on a background of space and time. Consequently, the physical laws must be adapted to any changes that might occur in the background (such as replacing the Galilean transformation by the Lorentz transformation); this scientific way of thinking about space and time, initiated by Einstein, led to modifications of the existing physical laws that are not Lorentz invariant \cite{Rindler}. Therefore, we can establish an elegant physical basis for MOND, if Milgrom's law Eq. (\ref{eq:mil}) is invariant under a new space-time coordinate transformation.

Let us consider a test particle of mass $m$ freely falling in a uniform gravitational field of a dark matter distribution, where the density distribution of dark matter is derived from the rotation curves of disk galaxies; the discrepancy between the rotation curve expected from the distribution of baryonic matter, $v_{B}^2=GM_B/r$, and the rotation curve measured by utilizing the Doppler effect, $v^2=GM/r$, yields the distribution of dark matter, $v_{D}^2=v^2-v_{B}^2=GM_D/r$. Then, the force acting on the test particle is given by Newton's second law
\begin{equation}\label{eq:newton2}
m\frac{d^2x}{dt^2}=mg_D=\frac{GmM_D}{r^2}.
\end{equation}
As a consequence of the equality of inertial and gravitational mass, a freely falling reference frame constitutes an inertial reference frame; the uniform gravitational field cannot be detected in the freely falling frame. We can demonstrate that the inertial mass of the test particle governed by the equation of motion Eq. (\ref{eq:newton2}) is equivalent to its gravitational mass by performing the space-time coordinate transformations,
\begin{equation}\label{eq:tr}
x'=x-\frac{1}{2}g_D t^2,
\end{equation}
\begin{equation}\label{eq:tr0}
t'=t.
\end{equation}
Where the spatial origins of the two coordinate systems $S$ and $S'$ coincide at $t'=t=0$, the unprimed system is the freely falling frame and the primed system is an inertial frame. Since the gravitational field $g_D$ is uniform (it does not depend on $t$ or $x$), the equation of motion becomes
\begin{equation}
m\frac{d^2x'}{dt'^2}=0,
\end{equation}
the gravitational force $mg_D$ is canceled by an inertial force. Hence, at any space-time point in a uniform gravitational field we can specify a locally inertial reference frame in accordance with the principle of equivalence. But, motivated by the existence of the acceleration scale Eq. (\ref{eq:max}), let us assume that at some space-time points in the uniform gravitational field $g_D$ we can not specify a locally inertial frame, in particular, let us postulate that there exists a universal constant of the order of the MOND acceleration constant, $a_{\dag} =\eta a_0$, which is invariant under transformations from inertial to accelerated frames. Thus, by performing a space-time coordinate transformations analogous to Eqs. (\ref{eq:tr}) and (\ref{eq:tr0}) when the magnitude of the gravitational field $g_D$ is equal to $a_{\dag}$, the equation of motion Eq. (\ref{eq:newton2}) must become
\begin{equation}
m\frac{d^2x'}{dt'^2}=ma_{\dag},
\end{equation}
this result is what we referred to earlier as the modified Newton's first law, an observer placed in the reference frame $S'$ is assumed to be isolated from the influence of any other matter; and yet the observer experiences a force of magnitude $ma_{\dag}$. This is due to the fact that $a_{\dag}$ is an invariant of coordinate transformations,
\begin{equation}\label{eq:newton3}
\frac{d^2x'}{dt'^2}=\frac{d^2x}{dt^2}=a_{\dag}.
\end{equation}
Then, according to our postulate, the space-time coordinate transformations Eqs. (\ref{eq:tr}) and (\ref{eq:tr0}) must be accommodated to the condition Eq. (\ref{eq:newton3}). Although this postulate has not been confirmed by any experiment and cannot be demonstrated from first principles, we will demonstrate that this postulate is the main reason for the emergence of the upper limit Eq. (\ref{eq:max}) that has been confirmed from observations.

Consider a uniformly accelerated reference frame $S'$ moving with an acceleration $g_D$ relative to an inertial reference frame $S$, if the origins of both reference frames coincide at $t'=t=0$ the origin of the reference frame $S'$ which has the coordinate $x'=0$ will be a distance $x=\frac{1}{2}g_Dt^2$ from the reference frame $S$. Thus, it is reasonable to assume that $x'$ is proportional to the same $(x-\frac{1}{2}g_D t^2)$ factor as in the familiar coordinate transformation Eq. (\ref{eq:tr}):
\begin{equation}\label{eq:tr1}
x'=\alpha (x-\frac{1}{2}g_D t^2),
\end{equation}
where $\alpha $ is the proportionality factor. The same argument applies if we take the coordinate system $S'$ to be the inertial frame, in this case, the origin of the reference frame $S$ has the coordinate $x=0$ and moves with acceleration $-g_D$ relative to the reference frame $S'$, so that $x'=-\frac{1}{2}g_Dt'^2$. Hence, the space transformation now takes the form
\begin{equation}\label{eq:tr2}
x=\alpha '(x'+\frac{1}{2}g_D t'^2),
\end{equation}
where $\alpha '$ is the proportionality factor. In order to determine the relation between the two factors $\alpha $ and $\alpha '$, let us consider two observers placed in the uniformly accelerated frames $S$ and $S'$. Since the observers in both frames experience the inertial force caused by the acceleration $g_D$ (only the sign of $g_D$ is different in the two frames), the uniformly accelerated frames $S$ and $S'$ must be equivalent for the description of physical events, for example, the length of the same measuring rod moving in these frames at the same acceleration $g_D$ must be the same.

Suppose a rod of length $l=x'_2-x'_1$ is at rest in the reference frame $S'$ which is moving with an acceleration $g_D$ relative to the reference frame $S$. The observer in $S$, who wants to measure the length of this rod, must measure the coordinates of the ends of the rod at the same time $t$. Using the coordinate transformation equation Eq. (\ref{eq:tr1}), we have
\begin{equation}
x'_1=\alpha (x_1-\frac{1}{2}g_D t^2), \quad x'_2=\alpha (x_2-\frac{1}{2}g_D t^2).
\end{equation}
Therefore, the length of the rod as measured in the reference frame $S$ is
\begin{equation}
\frac{l}{\alpha }=\frac{x'_2-x'_1}{\alpha }=x_2-x_1.
\end{equation}
Let us now interchange $S$ and $S'$. Suppose the same rod is at rest in $S$, where length $l=x_2-x_1$, the observer in $S'$, who wants to measure the length of this rod, must measure the coordinates of the ends of the rod at the same time $t'$. Using the coordinate transformation Eq. (\ref{eq:tr2}), we have
\begin{equation}
x_1=\alpha '(x'_1+\frac{1}{2}g_D t'^2), \quad x_2=\alpha '(x'_2+\frac{1}{2}g_D t'^2).
\end{equation}
Then, the length of the rod as measured in the reference frame $S'$ is
\begin{equation}
\frac{l}{\alpha '}=\frac{x_2-x_1}{\alpha '}=x'_2-x'_1,
\end{equation}
if both frames $S$ and $S'$ are equivalent and the length of the same rod moving in these frames at the same acceleration $g_D$ must be the same, we must have $l/\alpha =l/\alpha '$. Consequently,
\begin{equation}
\alpha =\alpha '.
\end{equation}

In accordance with the postulate Eq. (\ref{eq:newton3}), if there is an object moving at acceleration $a_{\dag}$ in an accelerating reference frame $S'$, then the trajectory of this object as measured by an observer in an inertial reference frame $S$ is
\begin{equation}
x=\frac{1}{2}a_{\dag} t^2,
\end{equation}
while the trajectory of the same object as measured by an observer in the accelerating frame $S'$ is
\begin{equation}
x'=\frac{1}{2}a_{\dag} t'^2.
\end{equation}
Substituting these trajectories into Eqs. (\ref{eq:tr1}) and (\ref{eq:tr2}), we obtain
\begin{equation}
a_{\dag} t'^2=\alpha t^2(a_{\dag}-g_D),
\end{equation}
\begin{equation}
a_{\dag} t^2=\alpha t'^2(a_{\dag}+g_D),
\end{equation}
from which we obtain a Lorentz-type factor
\begin{equation}\label{eq:tr3}
\alpha ^2 = \frac{a_{\dag}^2}{(a_{\dag}+g_D)(a_{\dag}-g_D)},
\end{equation}
\begin{equation}
\alpha = \frac{1}{\sqrt{1- g_{D}^2/a_{\dag}^2}}.
\end{equation}
The appearance of the Lorentz-type factor concludes our derivation of the maximum halo acceleration from physical assumptions, since it implies that it is a physical impossibility for a spherical dark matter halo to attain a surface density greater than $a_{\dag}G^{-1}$.

To get the time transformation, we can substitute Eq. (\ref{eq:tr1}) into Eq. (\ref{eq:tr2}) to obtain
\begin{equation}
x=\alpha (\alpha (x-\frac{1}{2}g_D t^2)+\frac{1}{2}g_D t'^2)
\end{equation}
and
\begin{equation}
\frac{1}{2}g_D t'^2=\frac{x}{\alpha }-\alpha (x-\frac{1}{2}g_D t^2)
                   = \alpha \frac{1}{2}g_D t^2 + (\frac{1}{\alpha }-\alpha ) x
\end{equation}
whereas, from Eq. (\ref{eq:tr3}), we have
\begin{equation}
\frac{1-\alpha ^2}{\alpha }= \sqrt{1- \frac{g_{D}^2}{a_{\dag}^2}} \frac{-(\frac{g_{D}^2}{a_{\dag}^2})}{(1-\frac{g_{D}^2}{a_{\dag}^2})}
                           = -\alpha \frac{g_{D}^2}{a_{\dag}^2},
\end{equation}
which leads to
\begin{equation}
\frac{1}{2}g_D t'^2=\alpha \frac{1}{2}g_D t^2 -\alpha \frac{g_{D}^2}{a_{\dag}^2} x.
\end{equation}
Thus, the space-time coordinate transformations Eqs. (\ref{eq:tr}) and (\ref{eq:tr0}) are the low acceleration limit $g_D\ll a_{\dag}$ of the Lorentz-type transformations
\begin{equation}\label{eq:tr4}
x'=\alpha (x-\frac{1}{2}g_D t^2),
\end{equation}
\begin{equation}\label{eq:tr5}
t'^2=\alpha (t^2-\frac{2g_D}{a_{\dag}^2}x).
\end{equation}

Unlike the Galilean and Lorentz transformations, the transformation Eqs. (\ref{eq:tr}) and (\ref{eq:tr0}), and the Lorentz-type transformation Eqs. (\ref{eq:tr4}) and (\ref{eq:tr5}) are nonlinear in the time coordinate. Hence, if we represent the nonlinear transformation for the space and time coordinates by a $4\times 4$ matrix
\begin{equation}\label{eq:matrix}
x'^\mu =\sum_{\nu =0}^{3} \Lambda _{\nu }^{\mu } x^\nu ,
\end{equation}
then the elements of the transformation matrix $\Lambda $ must be coordinate dependent $\Lambda (x)$. However, the new transformation Eqs. (\ref{eq:tr4}) and (\ref{eq:tr5}) has an advantage over the classical one Eqs. (\ref{eq:tr}) and (\ref{eq:tr0}): it has a free parameter with the dimensions of acceleration $a_{\dag}$. Since, $a_{\dag}$ is the same in all coordinate systems, it can be recognized (in analogy to the speed of light) as a conversion factor that converts time measurements in seconds to meters $t(meters)=\frac{1}{2}a_{\dag}\times t^2(seconds^2)$. Thus, we can define $x^0$ with the factor $a_{\dag}$ so that $x^0$ has the dimensions of length,
\begin{equation}
x^0=\frac{1}{2}a_{\dag} t^2
\end{equation}
and if we number the $x,y,z$ coordinates, so that
\begin{equation}
x^1=x, \quad x^2=y, \quad x^3=z
\end{equation}
then we can rewrite the Lorentz-type transformation Eqs. (\ref{eq:tr4}) and (\ref{eq:tr5}) in the four-vector notation:
\begin{equation}
x'^1=\alpha (x^1-\frac{g_D}{a_{\dag}}x^0), \quad x'^2=x^2, \quad x'^3=x^3
\end{equation}
\begin{equation}
x'^0=\alpha (x^0-\frac{g_D}{a_{\dag}}x^1)
\end{equation}
which we can write in the matrix form as Eq. (\ref{eq:matrix}), where the components of the transformation matrix are:
\begin{equation}
\Lambda _{\nu }^{\mu } = \left(
\begin{array}{cccc}
  \alpha  & -\alpha g_D/a_{\dag} & 0 & 0 \\
  -\alpha g_D/a_{\dag} & \alpha  & 0 & 0 \\
  0 & 0 & 1 & 0 \\
  0 & 0 & 0 & 1
\end{array}\right)
\end{equation}
Note that the components of the transformation matrix do not depend on the coordinates, as long as the gravitational field $g_D$ is uniform.

A four-vector can now be defined as any set of four components that transform under the Lorentz-type transformation Eq. (\ref{eq:matrix}) the same way $(x^0,x^1,x^2,x^3)$ do; for example, the time difference between any two events and their spatial separation can be represented by the displacement four-vector
\begin{equation}
dx^\mu =\left(
    \begin{array}{c}
      a_{\dag}tdt \\
      dx \\
      dy \\
      dz
      \end{array}
      \right)
\end{equation}
the components of this vector as specified relative to the coordinate system $S$ are related to the components of the same vector $dx'^\mu $ as specified relative to the coordinate system $S'$ by the Lorentz-type transformation Eq. (\ref{eq:matrix})
\begin{equation}\label{eq:matrix3}
dx'^\mu =\sum_{\nu =0}^{3} \Lambda _{\nu }^{\mu } dx^\nu .
\end{equation}
Let the inverse transformation to Eq. (\ref{eq:matrix3}) read as follows:
\begin{equation}\label{eq:matrix4}
dx^\nu =\sum_{\nu =0}^{3} \Lambda _{\beta }^{\nu } dx'^\beta ,
\end{equation}
where the matrices $\Lambda _{\nu }^{\mu }$ and $\Lambda _{\beta }^{\nu }$ are inverse to each other;
\begin{equation}
\Lambda _{\beta }^{\nu }= \left(
\begin{array}{cccc}
  \alpha  & \alpha g_D/a_{\dag} & 0 & 0 \\
  \alpha g_D/a_{\dag} & \alpha  & 0 & 0 \\
  0 & 0 & 1 & 0 \\
  0 & 0 & 0 & 1
\end{array}\right)
\end{equation}
By substituting Eq. (\ref{eq:matrix4}) into Eq. (\ref{eq:matrix3}), we can verify that the transformation matrix of the Lorentz-type transformation Eq. (\ref{eq:matrix}) satisfies the orthogonality condition
\begin{equation}
\sum_{\nu =0}^{3} \Lambda _{\nu }^{\mu } \Lambda _{\beta }^{\nu } =\delta _{\beta }^{\mu },
\end{equation}
where $\delta _{\beta }^{\mu }$ is the Kronecker delta.

Therefore, the scalar product of $dx^\mu $ with itself is an invariant quantity
\begin{equation}\label{eq:scalar}
dx^\mu dx_\mu =(a_{\dag}tdt)^2-(dx)^2-(dy)^2-(dz)^2,
\end{equation}
\begin{equation}
dx^\mu dx_\mu =(a_{\dag}t'dt')^2-(dx')^2-(dy')^2-(dz')^2.
\end{equation}
If an observer is at rest in the frame $S'$, then the spatial components of the displacement vector in this frame are zero
\begin{equation}
dx^\mu dx_\mu =(a_{\dag}tdt)^2-(dx)^2-(dy)^2-(dz)^2=(a_{\dag}t'dt')^2.
\end{equation}
Hence, the scalar product $dx^\mu dx_\mu $ is proportional to the time interval measured by an observer in its rest frame. We can employ this fact to define a transformation invariant coordinate time (the proper time $\tau $) which will allow us to obtain a four-vector when differentiating a four-vector,
\begin{equation}
dx^\mu dx_\mu =(a_{\dag}tdt)^2-(dx)^2-(dy)^2-(dz)^2=(a_{\dag}\tau d\tau )^2.
\end{equation}
Thus
\begin{equation}
\tau d\tau =tdt(1-\frac{1}{a_{\dag}^2}\frac{dx^2+dy^2+dz^2}{t^2dt^2})^{1/2}=tdt(1-g_{D}^2/a_{\dag}^2)^{1/2},
\end{equation}
the invariant time unit can be obtained by integrating both sides
\begin{equation}\label{eq:prop}
\tau =t(1-g_{D}^2/a_{\dag}^2)^{1/4}.
\end{equation}
The trajectory of the observer in the frame $S$ can be parameterized by the proper time $\tau $;
\begin{equation}\label{eq:coo}
x^\mu (\tau )=\left(
    \begin{array}{c}
      \frac{1}{2}a_{\dag} t^2(\tau ) \\
      \mathbf{x}(\tau )
      \end{array}
      \right)
\end{equation}
where $\mathbf{x}$ is the three-dimensional position, and therefore the excess acceleration four-vector can be obtained by differentiating the position four-vector Eq. (\ref{eq:coo}) with respect to the proper time Eq. (\ref{eq:prop})
\begin{equation}
\mathcal{A}^\mu =\frac{d^2x^\mu }{d\tau ^2}
=\alpha \left( 
         \begin{array}{c} 
           a_{\dag} \\
           d^2\mathbf{x}/dt^2
         \end{array}
         \right) =\alpha \left(
         \begin{array}{c} 
           a_{\dag} \\
           g_D
          \end{array}
          \right)
\end{equation}
Hence, we can generalize Newton's second law Eq. (\ref{eq:newton2}) to the covariant form
\begin{equation}\label{eq:for}
\mathcal{F}^\mu =m \mathcal{A}^\mu =m\alpha \left(
         \begin{array}{c} 
           a_{\dag} \\
           g_D
          \end{array}
          \right) =m\alpha \left(
          \begin{array}{c} 
           a_{\dag} \\
           GM_D/r^2
          \end{array}
          \right)
\end{equation}
where $\mathcal{F}^\mu $ is a four-vector force (the force due to the excess acceleration four-vector). We can find an appropriate interpretation of the time component of the force four-vector by Taylor expanding the $\alpha $ factor,
\begin{equation}\label{eq:time}
\mathcal{F}^0=\frac{ma_{\dag}}{\sqrt{1- g_{D}^2/a_{\dag}^2}}=ma_{\dag}+\frac{1}{2}m\frac{g_{D}^2}{a_{\dag}}+...
\end{equation}
let us now return to Milgrom's formula Eq. (\ref{eq:mil0}) and choose the following form of the interpolating function as chosen by Bekenstein \cite{bekenstein}
\begin{equation}
\mu (x)=\frac{\sqrt{1+4x}-1}{\sqrt{1+4x}+1},
\end{equation}
where $x=g/a_0$, hence, in the low acceleration limit $g\ll a_0$ the total acceleration due to Newtonian gravity can be expressed as follows
\begin{equation}
g=g_N+\sqrt{a_0 g_N}.
\end{equation}
Thus, the excess acceleration $g_D=g-g_N$ can take the following form
\begin{equation}
g_D=\sqrt{a_0 g_N}.
\end{equation}
Since in the equation Eq. (\ref{eq:time}) we can neglect the terms divided by $a_{\dag}^2$ and higher in the limit of small accelerations $g_D\ll a_{\dag}$, we can assume that $\mathcal{F}^0$ is made up of two parts: the first part gives identical results to Milgrom's law Eq. (\ref{eq:mil}) in the low acceleration limit (when the numerical factor takes the value $\eta =\frac{1}{2}$) and in the presence of Newtonian gravitational forces
\begin{equation}\label{eq:motion}
\mathcal{F}^{0}_1 =\frac{ma_{\dag}}{\sqrt{1- g_{D}^2/a_{\dag}^2}} - ma_{\dag}=mg_N
\end{equation}
and the second part is a constant force
\begin{equation}
\mathcal{F}^{0}_2 =ma_{\dag}
\end{equation}
which is the force experienced by an observer at rest, and it can be interpreted as a location independent weight, in contrast, the weight of an object in Newtonian physics is defined as the product of the object's mass and the magnitude of the gravitational acceleration which depends on the location. We deduce from this that $\mathcal{F}^0=\mathcal{F}^{0}_1 +\mathcal{F}^{0}_2 =F$ is the total force acting on the particle mass $m$.

In the Newtonian limit $g\gg a_0$ or equivalently $a_{\dag}\rightarrow \infty $, i.e. $\alpha =1$, the spatial components of the force four-vector Eq. (\ref{eq:for}) reduce to Newton's second law Eq. (\ref{eq:newton2}). Note that there is no need for a predefined interpolation function, but instead, it is the Lorentz-type factor $\alpha $ that allows the transition between the Newtonian and MONDian regime.

In order to write the equation of motion Eq. (\ref{eq:for}) using the Lagrangian formalism, the classical Lagrangian $L=\frac{1}{2}m\mathbf{u}^2-V(\mathbf{x})$ must be invariant under the Lorentz-type transformations. The Lagrangian must be a function of the coordinates Eq. (\ref{eq:coo}) and their derivatives with respect to the invariant parameter -the proper time $\tau $-. Suppose the force Eq. (\ref{eq:for}) is a conservative force derivable from a potential,
\begin{equation}
\mathcal{F}^\mu =-\frac{\partial V(x^\mu )}{\partial x^\mu }
\end{equation}
Using the velocity four-vector
\begin{equation}
u^\mu =\frac{dx^\mu }{d\tau }=\alpha  ^{\frac{1}{2}} \left(
    \begin{array}{c}
      a_{\dag} t \\
      \mathbf{u}
      \end{array}
      \right)
\end{equation}
we can suggest the covariant Lagrangian
\begin{equation}\label{eq:lag}
L =\frac{1}{2}mu^\mu u_\mu - V(x^\mu )
\end{equation}
It follows from Hamilton's variational principle
\begin{equation}
\delta S =\delta \int_a^b L \, d\tau =0
\end{equation}
that the Lagrangian Eq. (\ref{eq:lag}) must satisfy the Lagrange equations
\begin{equation}
\frac{d}{d\tau }(\frac{\partial L}{\partial u^\mu })-\frac{\partial L}{\partial x^\mu }=0
\end{equation}
Since the MOND effects can be attributed to the presence of a fictitious dark halo, the time component of Lagrange's equations determines the distribution of the dark halo from the distribution of baryonic mass
\begin{equation}
\frac{d}{d\tau }(\frac{\partial L}{\partial u^0})=\frac{ma_{\dag}}{\sqrt{1- g_{D}^2/a_{\dag}^2}}
\end{equation}
\begin{equation}
\frac{\partial L}{\partial x^0}=-\frac{\partial V}{\partial x^0}=mg_N+ma_{\dag}
\end{equation}
where $\mathcal{F}^0=mg_N+ma_{\dag}$ is the total force exerted on the particle. While the spatial components of Lagrange's equations determine the motion of test particles in the gravitational field of the dark halo
\begin{equation}
\frac{d}{d\tau }(\frac{\partial L}{\partial u^i})=m\alpha \frac{d\mathbf{u}}{dt}
\end{equation}
\begin{equation}
\frac{\partial L}{\partial x^i}=-\frac{\partial V}{\partial x^i}=m\alpha g_D
\end{equation}

The formulation of MOND, illustrated above, not only reproduces the predictions of Milgrom's law but it also leads to a number of physical consequences that arise from replacing the space-time coordinate transformation Eqs. (\ref{eq:tr}) and (\ref{eq:tr0}) by the Lorentz-type transformation Eqs. (\ref{eq:tr4}) and (\ref{eq:tr5}).

Let us first write the transformations Eqs. (\ref{eq:tr4}) and (\ref{eq:tr5}) in the differential form
\begin{equation}\label{eq:lorentz4}
dx'=\alpha (dx-g_D tdt),
\end{equation}
\begin{equation}\label{eq:lorentz5}
t'dt'=\alpha (tdt-\frac{g_D}{a_{\dag}^2}dx).
\end{equation}
Consider a rod of length $dx'$ placed at rest in a frame of reference $S'$ which is moving relative to a frame of reference $S$ with an acceleration of $g_D$. To measure the rod's length in the frame $S$, the end points of the rod must be observed at the same time $t$. Since the observer in the frame $S$ must measure the distance between the two end points simultaneously $dt=0$, we have from Eq. (\ref{eq:lorentz4}),
\begin{equation}
dx=(1- g_{D}^2/a_{\dag}^2)^{1/2} dx'.
\end{equation}
If, for example, $D$ is the distance between two stars in a binary pair whose positions are observed simultaneously, then
\begin{equation}
D=(1- g_{D}^2/a_{\dag}^2)^{1/2} D_0,
\end{equation}
where $D_0$ is the distance between the two stars as measured in a frame of reference in which $g_D$ is equal to zero. Therefore, the distance between two uniformly accelerating stars with an acceleration of $g_D$ is reduced by a factor $(1- g_{D}^2/a_{\dag}^2)^{1/2}$.

Consider a clock placed at rest in a frame of reference $S'$ and it measures a time interval $dt'$, suppose that $S'$ is moving relative to a frame of reference $S$ with an acceleration of $g_D$. Since the clock is stationary (there is no spatial displacement $dx'=0$) in the frame $S'$, we have from Eq. (\ref{eq:lorentz4}),
\begin{equation}\label{eq:lorentz6}
dx'=\alpha (dx-g_D tdt)=0
\end{equation}
Substituting Eq. (\ref{eq:lorentz6}) into Eq. (\ref{eq:lorentz5}) we obtain
\begin{equation}\label{eq:lorentz7}
t=\frac{t'}{(1- g_{D}^2/a_{\dag}^2)^{1/4}}
\end{equation}
we conclude from the above formula that a uniformly accelerating clock with an acceleration of $g_D$ runs slow by a factor $(1- g_{D}^2/a_{\dag}^2)^{1/4}$ relative to the clocks in the frame $S$.

Radiation emitted from a source while moving directly toward a receiver will be shifted in frequency by the Doppler effect. Suppose a light source is at rest in a frame of reference $S'$ which is moving toward a receiver in a frame of reference $S$ with speed $u$, thus if the light source sends a signal at a time interval $dt$ as measured by a co-moving observer at the source, then during that time the signal is sent from position $x=udt$, so this signal arrives at the receiver time
\begin{equation}\label{eq:doppler}
dT=dt-dtu/c
\end{equation}
apart. But the effect of time dilation Eq. (\ref{eq:lorentz7}) modifies Eq. (\ref{eq:doppler}) to
\begin{equation}
dT=\frac{dt'-dt'u/c}{(1- g_{D}^2/a_{\dag}^2)^{1/4}},
\end{equation}
where $dt'$ and $dT$ are inversely proportional to the frequency $\nu _0$ of the source in $S'$ and the frequency $\nu $ of the source as seen by the observer in $S$, respectively,
\begin{equation}\label{eq:doppler2}
\frac{\nu _0}{\nu }=\frac{1-u/c}{(1- g_{D}^2/a_{\dag}^2)^{1/4}}.
\end{equation}
Therefore, if we consider a spectroscopic binary star system placed in a frame of reference in which $g_D$ is not equal to zero, then Eq. (\ref{eq:doppler2}) predicts that there should be a detectable Doppler shift at any given time even when the inclination of the star's orbit relative to the line of sight is zero, i.e. the radial component of the star's velocity is zero $u=0$. In contrast, the classical Doppler relation for non-relativistic speeds $\nu _0=\nu (1-u/c)$ predicts that there will be no detectable Doppler shift at the instants of time when $u=0$.

\section{MOND and clusters of galaxies}

We shall illustrate how the weak equivalence principle or the universality of free fall which allows us to equate the Newtonian gravitational field with an accelerated reference frame can be incorporated into our formulation of MOND. In Newtonian mechanics, an object freely falling in a uniform gravitational field is considered to be weightless, however, it seems from the modified Newton's first law $F=m\eta a_0$ that the state of weightlessness cannot actually be achieved even if the object is in a state of free fall in a uniform gravitational field. So combining the universality of free fall with the modified Newton's first law yields the following empirical formula that replaces Milgrom's formula,
\begin{equation}\label{eq:new}
g\mu (g/a_0)=g_N+\eta a_0
\end{equation}
where the interpolating function can be chosen to resemble the $\mu $-function of Milgrom's formula, such that it satisfies $\mu (x)=1$ when $x\gg \eta $, and $\mu (x)=x$ when $x\ll \eta $. The dimensionless constant $\eta $ is equal to $1/2$ according to the argument above Eq. \ref{eq:motion}.

Then, in high acceleration systems $g\gg \frac{1}{2}a_0$ the term $\frac{1}{2}a_0$ appears as an anomalous acceleration,
\begin{equation}
g=g_N+\frac{1}{2}a_0
\end{equation}
while in the low acceleration limit $g\ll \frac{1}{2}a_0$ we obtain
\begin{equation}\label{eq:mo}
g=\sqrt{a_0 g_N +\frac{1}{2}a_{0}^2}
\end{equation}
which might becomes relevant within large clusters of galaxies, particularly within their central regions, since MOND fails to completely resolve the mass discrepancy problem in these systems \cite{Sander} \cite{Aguirre}. We can, in principle, demonstrate this by considering a cluster in hydrostatic equilibrium, using Eq. (\ref{eq:mo}) the dynamical mass can be determined from the density and temperature distribution of the X-ray emitting gas,
\begin{equation}
\sqrt{a_0 GM +\frac{1}{2}a_{0}^2 r^2}=-\frac{kT}{\mu m_p}(\frac{d\ln \rho }{d\ln r}+\frac{d\ln T}{d\ln r})
\end{equation}
This relation is apparently more convenient for clusters than the mass-temperature relation $M\propto T^2$ predicted by MOND \cite{Aguirre}, because clusters are mostly isothermal, and isothermality (in the case of the mass-temperature relation) corresponds to a point mass not to an extended object.

Even though, the formula Eq. (\ref{eq:new}) seems to be helpful in removing the remaining mass discrepancy in MOND, it is not obvious how this formula is consistent with the fact that rotation curves are asymptotically flat. The resolution of this apparent contradiction lies in the principle of equivalence; since the ratio of inertial to gravitational mass is the same for all bodies $m_i=m_g$ then it should be possible, in the case of uniform gravitational field, to transform to space-time coordinates such that the effect of a gravitational force will not appear
\begin{equation}
m_i g'\mu (g'/a_0)=(m_g-m_i)g_N+m_i\eta a_0=m_i \eta a_0
\end{equation}
Hence, the analogue of Einstein's principle of equivalence in MOND states that: it is possible, in a sufficiently small regions of space-time such that the Newtonian gravitational field changes very little throughout it, to specify a coordinate system in which matter satisfies the law of motion Eq. (\ref{eq:for}), and hence it is possible in these regions to observe the asymptotic flatness of rotation curves. Therefore, any consistent generalization of the afore mentioned Lorentz-type invariance to non-uniformly accelerated coordinate systems must reproduce the results of the formula Eq. (\ref{eq:new}).

\section{Conclusion}

In this paper, we have presented a relativistic formulation of the MOND hypothesis based on the assumption that accelerated measuring rods and clocks are affected by acceleration in the low acceleration regime. The proposed relativistic formulation produces a prediction that does not result from either the MOND hypothesis or the dark matter hypothesis; it predicts that spectroscopic binary star systems in the low acceleration $g\ll a_0$ and low velocity $u\ll c$ regime should exhibit a non-classical Doppler shift, as expressed by the formula Eq. (\ref{eq:doppler2}), due to a time dilation effect. We also showed that the mass discrepancy in clusters of galaxies can be accounted for by a consistent generalization of the Lorentz-type symmetry to non-uniformly accelerated coordinate systems.

\section*{Acknowledgement}
I thank Stacy McGaugh and the anonymous referees for helpful comments. The endorsement of Pavel Kroupa
to submit this paper to the arXiv is gratefully acknowledged.



\end{document}